\documentstyle[preprint,aps,tighten]{revtex}
\newcommand{\be}{\begin{equation}}
\newcommand{\ee}{\end{equation}}
\newcommand{\br}{\begin{eqnarray}}
\newcommand{\er}{\end{eqnarray}}

\newcommand{\half}{\frac{1}{2}}
\def\a{{\alpha}}
\def\b{{\beta}}
\def\g{{\gamma }}

\def\m{{\mu}}
\def\n{{\nu}}

\begin{document}
\draft
\title{ Patterns of Broken Chiral Symmetry in 
Quantum Chromodynamics  }
\author{R. Acharya $^a$ and P. Narayana Swamy $^b$}
\address{$^a$ Physics Department, Arizona State University, Tempe, AZ 85287
\\
$^b$ Physics  Department, Southern Illinois University,
Edwardsville, IL 62026}
\maketitle

\begin{abstract}

We establish that in Quantum Chromodynamics (QCD) at zero temperature, $SU_{L+R}(N_F)$ exhibits the vector mode conjectured by Georgi  and $SU_{L-R}(N_F)$ is realized in  either the Nambu-Goldstone mode or else $Q_5^a$ is also screened from view at infinity. The Wigner-Weyl mode is ruled out unless the beta function in QCD develops an infrared stable zero.

\end{abstract}

\vspace{3.1in}
\noindent October 18, 2000\\
\noindent This is a revised version of the work previously reported in arXiv:hep-ph/9912389.\\
\noindent ($^a $): electronic address: raghunath.acharya@asu.edu\\
\noindent ($^b $): electronic address: pswamy@siue.edu\\
\noindent 

We shall begin by summarizing the key results constituting  conventional wisdom in 
standard Quantum Chromodynamics (QCD). 

Firstly, in vector-like gauge theories and in QCD in particular, non-chiral 
symmetries  such as $SU_{L+R}(2) \subset SU_L(2) \times SU_R(2)$  or $SU_{L+R} 
(3)
\subset  SU_L(3) \times SU_R(3)$ {\it cannot\/} be spontaneously broken. This 
is the Vafa-Witten result \cite{vafa,rapn1}.  In QCD with large $N_c$, 
Coleman and Witten \cite{CW} showed that chiral symmetry is broken to diagonal $U(N_F)$ 
and thus, if chiral symmetry is broken, it must happen in such a manner that 
flavor symmetry is preserved. Secondly, chiral $SU_L(3) 
\times SU_R(3)$ symmetry  in QCD with massless $u, d, s$ quarks must be spontaneously 
broken. However,  it is difficult to show \cite{QFT} that chiral $SU_L(2) \times SU_R(2) 
$ symmetry in QCD with massless $u, d$ quarks  is also spontaneously broken. 
Thirdly, QCD may very well  exhibit the Higgs mode for the vector current and 
the Goldstone mode for the axial vector current {\it i.e., \/} the massless 
scalars arising from Goldstone theorem get `eaten up' by the gauge vector 
field which consequently acquires a finite mass. This important conjecture was 
introduced by Georgi in 1989 who called it as a new realization of chiral 
symmetry (``vector mode" ) since it involves both the Wigner-Weyl and 
Nambu-Goldstone modes.  This is Georgi's conjecture \cite{Georgi}. 

To quote Weinberg \cite{Salam}, ``A recent paper of Georgi, can be 
interpreted as proposing that QCD at zero temperature is near a second order 
phase transition, at which the broken chiral $SU(3) \times SU(3)$ symmetry has 
a $(8,1) + (1,8)$ representation, consisting of the octet of pseudoscalar 
Goldstone bosons plus an octet of massless scalars, that on the broken symmetry 
side of the phase transition, become the helicity-zero states of the massive 
vector meson octet,  $\cdots$.   It is intriguing and mysterious that at the second 
order phase transition at which  chiral $SU(2) \times SU(2)$ of massless 
QCD becomes {\it unbroken \/}, this symmetry may become {\it local\/} with $\rho$ and $A_1$ as 
massless gauge bosons".

As a 
final point one may observe that the charges corresponding to spontaneously 
broken local gauge symmetries are {\it screened\/} and the vector mesons are 
massive. This is a manifestation of spontaneously broken local symmetries. For 
instance, the well-known example \cite{Higgs} is the Abelian Higgs model: in 
the spontaneously broken phase, the vector field has a finite mass (thus the 
field is of finite range) and the conserved current does not have a total 
charge in the physical Hilbert space. 

Coleman and Witten \cite{CW} have demonstrated that QCD exhibits spontaneously 
broken chiral symmetry by examining the large $N_c$ limit. In spite of the 
existence of a great deal of work on this subject using several such 
approaches \cite{tHooft}, what is lacking is a general rigorous proof of spontaneously 
broken chiral symmetry in QCD.  

The nature of spontaneously broken chiral symmetry is intimately connected to 
spontaneously broken scale invariance. It has been emphasized by Adler 
\cite{AdlerRMP}  that there are two examples of relativistic field theories 
which exhibit spontaneously broken scale invariance where chiral symmetry is 
also broken. These are Johnson-Baker-Wiley model of quantum 
electrodynamics \cite{BJW} and asymptotically free gauge theories. This indeed may be a 
general feature as we pointed out recently in our investigation \cite{rapn2} 
of spontaneously broken chiral symmetry in QCD. Let us review this connection 
briefly.

Unbroken scale invariance can be expressed as
\be
Q_D(t) | 0> \;  = 0,
\label{1}
\ee
where the dilatation charge is 
\be
Q_D(t)= \int \; d^3x \; D_0( {\bf x}, t),
\label{2}
\ee
defined in terms of $D_{\m}( {\bf x},t )$, the dilatation current. 
Equivalently,
\be
\partial^{\m} D_{\m} |0>=0.
\label{3}
\ee
Invoking Coleman's theorem    \cite{Coleman}, which is valid for 
continuous symmetries,  we can then prove that the divergence of the 
dilatation current itself must itself vanish identically:
\be
\partial^{\m} D_{\m}=0.
\label{4}
\ee 
On the other hand, it turns out that we know the divergence of the dilatation current as it is 
determined by the trace anomaly \cite{Collinsetal}  in QCD:
\be
\partial^{\m}D_{\m}= \half \; \frac{\b(g)}{g} G^{\a}_{\m \n} G^{\m \n}_{\a} + 
\sum_i \; m_i [1 + \g_i (\theta)] \bar{\psi}_i  \psi_i,
\label{5}
\ee
where the second term vanishes for massless quarks in the chiral limit. 
Consequently the beta function must vanish. It is well-known that in an asymptotically free theory 
of QCD which also exhibits confinement, the behavior of $\b(g)$ is such that 
it decreases as $g$ increases and never turns over. Consequently $g=0$ is the 
only possibility and hence the theory reduces to triviality. We therefore 
conclude by {\it reductio ad absurdum\/} that scale invariance must be broken 
spontaneously by the QCD vacuum state
\be
Q_D(t)| 0>  \; \not= 0.
\label{6}
\ee
It is important to note that since $g \not =0$ in the standard theory of QCD, 
scale invariance is also explicitly broken by the trace anomaly.  In other 
words, scale invariance is broken both spontaneously by the vacuum state and 
explicitly by the trace anomaly.

We shall establish that QCD at zero temperature, satisfying both asymptotic freedom and confinement, exhibits the following features: 
(a) $SU_{L+R}(N_F)$  exhibits the vector mode conjectured by Georgi  \cite{Georgi} (Georgi-Goldstone mode). (b) $SU_{L-R}(N_F)$  exhibits 
either the Nambu-Goldstone mode or else the axial-vector charge $Q_5^a$ is also screened from view at infinity. If the latter case were to occur, then QCD confines without breaking chiral symmetry: both $ SU_{L+R}(N_F)$ and $SU_{L+R}(N_F)$ are realized in the Higgs mode (Georgi-Wigner mode), with no scalar or pseudoscalar Nambu-Goldstone bosons and the vector and axial-vector mesons are degenerate. (c) The Wigner-Weyl mode corresponding  to $Q^a |0 \rangle  =0, \;   Q_5^a |0 \rangle =0$  is ruled out: the Callan-Symanzik beta function has to turn over to yield an infrared stable fixed point at a finite value of $g$ if chiral symmetry is to be restored. 

We now sketch the proof of these interesting assertions.  We begin with the vector current $V_{\m}^a$ and its conservation 
\be
\partial^{\m}V_{\m}^a ({\bf x},t) =0 \; .
\label{7}
 \ee
This  implies the local version
\be
[Q^a (t), \; H({\bf x}, t) ] =0
\label{8}
\ee
where $H({\bf x},t ) = \Theta^{00}$ is the Hamiltonian density, if  
 the surface terms at infinity can be discarded. This is clearly  justified  if the flavor vector charge annihilates the vacuum, 
\be
Q^a(t) |0 \rangle =0
\label{9}
\ee
which is guaranteed by the Vafa-Witten theorem \cite{vafa,rapn1} {\it i.e., } non-chiral symmetries cannot be spontaneously broken in vector-like gauge theory. Hence there are no scalar Nambu-Goldstone bosons to produce a long range interaction, which in turn would have  resulted in a non-vanishing contribution to the surface terms.

Let the commutator
\be
[Q_D (0),Q^a (0) ]= -i d_Q Q^a (0)
\label{10}
\ee
define the scale dimension $d_Q$ of the charge $Q^a(0)$. By translation invariance, this can be put in the form
\be
[Q_D (t),Q^a (t) ]= -i d_Q Q^a (t).
\label{11}
\ee
It is important to stress that operator relations such as the above equation are unaffected by spontaneous symmetry breaking  as emphasized by Weinberg  \cite{QFT}.  Let us consider  the double commutator which follows from Eq.(\ref{8}),
\be
[Q_D(t), \, [Q^a(t),\, H]\, ]=0,
\label{12}
\ee
where $Q_D(t)$ is the dilation charge defined in Eq.(\ref{2}). If we now invoke the Jacobi identity we can recast the above equation in the form
\be
\left  [Q^a(t), \, [H,\, Q_D(t)\, ]\, \right ] + \left  [H, [ Q_D(t), Q^a(t)] \, \right ] =0.
\label{13}
\ee
Since
\be
[H ( {\bf x},t), \, Q_D(t)] = -i \partial_{\m }D^{\m} ( {\bf x},t) \not=0,
\label{14}
\ee
by virtue of the trace anomaly, Eq.(\ref{5}), and  making use of 
Eqs.(\ref{11}, \ref{8}), we arrive at the  {\it operator} relation
\be
[Q^a(t), \, \partial^{\m}D_{\m} ( {\bf x}, t ) \, ] =0 \, .
\label{15}
\ee
Applying  this relation on  the vacuum state, we obtain
\be
[Q^a(t), \, \partial^{\m}D_{\m} ( {\bf x}, t ) \, ] |0 \rangle =0 \, .
\label{16}
\ee
We may now invoke the result of Vafa-Witten theorem \cite{vafa} 
\be
Q^a(t) |0 \rangle=0\; ,
\label{17}
\ee
and conclude that
\be
{\cal O}( {\bf x}, t) | 0 \rangle \equiv Q^a(t) \partial^{\m}D_{\m} ( {\bf x }, t ) \,  |0 \rangle =0 \, .
\label{18}
\ee

The divergence of the dilatation curent is clearly a local operator. Since the vector current $J_{\m}^a$ is conserved,  the vector charges $Q^a$ are time-independent and they have the following important properties: (a) They are constants of motion ({\it i.e.,} constant operators)  and (b) they are also the generators of the vector algebra. It is worthwhile elaborating this point. Let us for instance, consider the vector charge $Q^a$ for $a=3$. Since the vector meson $\rho$ is coupled to a conserved vector current $J_{\m}^a$ , where $a=1,2,3$, (this is mandatory for a massless vector meson and is valid for a massive vector meson in the Landau gauge),  then one can demonstrate, following Kroll, Lee and Zumino \cite{Kroll}, by invoking $T, C$ and baryon number invariance of the theory,  that one must have $\int \; J_0^3 ( {\bf r},0) d^3r = \lambda T_3$, where $\lambda$ is a constant and $T_3$ is the third component of the  $SU_V(2)$ generator. If any of the quark masses were to vanish, then the theta angle would have no effect and there would be no $P$ or $CP$ nonconservation in QCD \cite{QFT}. 
 In other words, Eq.(\ref{18}) for $a=3$ becomes
\be
\lambda T_3\; \partial^{\m}D_{\m} |0 \rangle = 0\; .
\label{19}
\ee
We can now invoke the Federbush-Johnson theorem and arrive at the result
$\lambda T_3 \partial^{\m}D_{\m} = 0\; $.

Since $ \partial^{\m}D_{\m} ( {\bf x }, t ) $ cannot vanish 
in a theory of QCD which exhibits both asymptotic freedom and confinement  except at $g=0$, we are led to conclude that  $\lambda = 0$ {\it i.e., } the third component of the vector charge, $Q^3$ is screened: $Q^3 \equiv 0$. By invoking $SU_V(3)$ invariance, this argument generalizes to 
$Q^a \equiv 0$ where $a=1, 2, \cdots 8$ and thus we arrive at the conclusion that the vector charge must be screened. Hence by the Higgs mechanism, the vector mesons become massive. {\it It is worth stressing that the application of Federbush-Johnson theorem goes through since the locality of $\partial^{\m}D_{\m}$ remains undisturbed upon multiplication by a constant operator. The important result, $Q^a =0$,  is a manifestation of spontaneously broken local symmetry }. The vector mesons become massive and the scalar would-be Nambu-Goldstone bosons disappear.

Let us now consider the axial-vector charges $
Q_5^a$.  The Vafa-Witten theorem does not apply in this case and hence we proceed by the method of {\it reductio ad absurdum} as follows. We assume
 $ Q_5^a | 0\rangle =0$ corresponding to the Wigner-Weyl mode of unbroken symmetry and $Q_5^a$ are conserved and time independent. 
We begin by defining the scale dimension of the axial-vector charge by
\be
[Q_D (0),Q_5^a  ] = -i d_{Q5} Q_5^a \;  .
\label{20}
\ee
Repeating the earlier analysis now for the axial-vector charges, exactly as in Eqs.(\ref{11} -- \ref{18}), we obtain
\be
Q_5^a \, \partial^{\m}D_{\m} ( {\bf x}, t )  | 0 \rangle = 0 \, .
\label{21}
\ee
Since $Q_5^a $  is a non-zero constant for the assumed Wigner-Weyl mode, applying the Federbush-Johnson theorem, we arrive at the conclusion
\be
Q_5^a \, \partial^{\m}D_{\m} ( {\bf x}, t )  \equiv 0 \, .
\label{22}
\ee
This in turn requires  $\partial^{\m}D_{\m} ( {\bf x}, t )=0$. This would imply that QCD exhibiting both asymptotic freedom and confinement is a free field theory. Hence by {\it reductio ad absurdum} we are led to the conclusion: either $Q_5^a |0 \rangle \not=0$ or $Q_5^a \equiv 0$. The first alternative is the Nambu-Goldstone realization of chiral symmetry which must hold \cite{QFT} for $N_F=3 $. The second alternative in conjunction with the screening of the vector charges, {\it  i.e.,} $Q^a \equiv 0, \, Q_5^a \equiv 0\; $,  is the Higgs mode alternative (Georgi-Wigner mode): both scalar and pseudoscalar Nambu-Goldstone bosons have been devoured. This case corresponds to confinement with exact chiral symmetry.  Such a mode is realized in Supersymmetric QCD \cite{Intr}.

We note that the Kroll-Lee-Zumino argument used above for the vector charge is not valid for the case of the axial-vector charge since the axial baryon number invariance is violated by $U_A(1)$ anomaly \cite{QFT}. This is consistent with the fact that no algebra for the axial-vector charge is assumed. 

In conclusion, it is interesting that the Wigner-Weyl mode is ruled out in QCD at $T=0$, if both asymptotic freedom and confinement obtains. This follows from Eqs.(\ref{19},\ref{22}). Hence the Wigner-Weyl mode can occur only if the beta function turns over to yield an infrared stable fixed point \cite{Appel}. 

An effective Lagrangian  \cite{bando,sannino} which realizes the Georgi-Goldstone mode is easily constructed, following the general procedure for building models with vector and axial vector mesons.

Finally, it is important to stress that we have not invoked the $SU_L(N_F) \times SU_R(N_F)$ charge algebra in our analysis. In view of the screened charges, {\it i.e.,} $Q^a =0, \; Q^a_5|0\rangle \not= 0 $ for the Georgi-Goldstone mode and $ Q^a= Q^a_5=0$ for the Georgi-Wigner mode, one must revert back to local current algebra \cite{QFT}which leads to the Weinberg sum rules in QCD. There still remain some unresolved issues that need to be addressed. First, is the vacuum expectation value of ${\bar q}\, q$  ``small" near the vector limit? If so, how does one explain the lattice \cite{lattice} results? Secondly, what is the precise criterion which signals the distinction between the Georgi-Goldstone and the Georgi-Wigner modes? Finally in QCD the conserved vector charges for $N_F=2$ are just isospin operators. Therefore what is the real meaning behind their vanishing? These and other related issues will be addressed in a forthcoming work.

\acknowledgments

We are grateful to Professor S. Adler for drawing our attention to the work of Intrilligator and Seiberg,  and  Seiberg and Witten; in particular to the phase of confinement without chiral breaking in supersymmetric QCD for $N_F = N_C +1\, $.



\begin{thebibliography}{30}

\bibitem {vafa} C. Vafa and E. Witten 1984 {\it Nucl. Phys. \/} {\bf B 234}, 
 173; {\it ibid} 1984 {\it Comm. Math. Phys.\/} {\bf 95}  257.

\bibitem {rapn1}  R. Acharya and P. Narayana Swamy 1987 {\it Nuovo Cimento\/} 
{\bf A 98}  773; {\it ibid} 1989 {\it Nuovo Cimento\/} {\bf A 101 }  607.

\bibitem {CW}  S.Coleman and E.Witten 1980 {\it Phys.  Rev.Lett.}  {\bf 
45}  100.

\bibitem {QFT}  S.  Weinberg 1996  {\it  The  Quantum  Theory  of 
Fields\/},   Volume  II  (Cambridge:  Cambridge  University   Press.  See also S.Weinberg 1990 {\it Phys. Rev. Lett.} {\bf 65},   1177.

\bibitem {Georgi} H. Georgi 1989 {\it Phys. Rev. Lett.\/} {\bf 63 }  1917; {\it ibid } 1990 {\it Nucl. Phys. \/} {\bf B 331}  311--330.

\bibitem {Salam} S. Weinberg 1994 in {\it Salam Festschrift, ICTP Trieste, 
Italy, March 1993\/}, Editors A. Ali {\it et al\/}  World Scientific Series in 
20th century physics, Vol. 4.

\bibitem {Higgs} P. W. Higgs 1966 {\it Phys. Rev.\/} {\bf 145}  1156; G. 
Guralnik, C.R. Hagen and T.W.B. Kibble 1964 {\it Phys. Rev. Lett.\/} {\bf B13}, 
 585. The non-Abelian case is treated by T.W.B.Kibble 1967 {\it Phys. 
Rev.\/} {\bf 155}  1554.

\bibitem {tHooft} G.'t Hooft 1980 in {\it Recent Developments  in 
Gauge theories\/}, editors G.t'Hooft {\it et al\/} (New  York: Plenum Press 
) reprinted in {\it  Dynamical  Gauge  Symmetry 
Breaking\/} 1982  edited  by E.Farhi and R.Jackiw (Singapore: World  Scientific 
Publishing  Co.); T.Appelquist,  J.Terning  and 
L.Wijewardhana 1996 {\it Phys.  Rev.  Lett.} {\bf  77}   1214;  J. 
Cornwall 1980  {\it Phys.  Rev. } {\bf D22}  1452; A.  Casher 1979 {\it Phys. 
Lett.}  {\bf  83B},  395; O. Nachtman  and  W.Wetzel 1979 {\it Phys. 
Lett.}  {\bf  81B} 211; D.Gross and A. Neveu 1974  {\it Phys.  Rev. }
{\bf  D10} 3235; K. Lane 1974 {\it Phys. Rev.}  {\bf  D10},  1353. See also R. Acharya and P. Narayana Swamy 1986 in {\it  A.I.P 
Conference \\
Proceedings  \/} {\bf 150}, ed., D. Geesaman, Lake  Louise 
(New York: American Institute of Physics);  C.Callan, R.Dashen 
and D.Gross 1978  {\it Phys. Rev.} {\bf D17}  2717.

\bibitem {AdlerRMP} S. Adler 1982 {\it Rev. Mod. Phys.} {\bf  54}   729.

\bibitem {BJW} K. Johnson, M. Baker and R. Willey 1964 {\it Phys.  Rev. }
{\bf 136},  1111B.

\bibitem {rapn2} R. Acharya and P. Narayana Swamy 1997 {\it Mod. Phys. Lett.\/} 
{\bf A 12}  1649-1654. The analysis in this earlier work assumed the 
existence of $SU_L(N) \times SU_R(N)$ charge algebra which implies the 
canonical values $d_Q = d_{Q_5}=0$. In the present work the constraint of the 
charge algebra is not imposed.

\bibitem {Coleman}  S.Coleman 1966 {\it J. Math. Phys.}  {\bf  7}   787. See also  R. Streater and A.Wightman 1964 {\it PCT, Spin  \& 
Statistics, and all that\/} (New york: W. Benjamin, Inc.); P. Federbush 
and K. Johnson 1960  {\it Phys.  Rev. }
{\bf 120}  1926  F. Strocchi 1972 {\it Phys. Rev.} {\bf  D6}   1193.  This  work  extends  the  Federbush-Johnson  theorem  to 
theories  with  indefinite  metric.  Local  gauge  quantum  field 
theories require an indefinite metric. See F. Strocchi 1978 {\it Phys.Rev.} 
{\bf D17}  2010. 

\bibitem {Collinsetal}  S. Adler, J. Collins and  A.  Duncan 1977
{\it Phys.  Rev.}  {\bf D15}  1712; J. Collins,  A.  Duncan,  S. 
Joglekar 1977 {\it  Phys. Rev.} {\bf D16}  438.

\bibitem{Kroll}
N. Kroll, T.D.Lee and B. Zumino 1967 {\it Phys. Rev.} {\bf 157}  1376. See also J. Bernstein 1968 {\it Elementary Particles and their Currents} (San Francisco: W.H. Freeman and Company).

\bibitem{Intr}
K. Intrilligator and N. Seiberg 1996 {\it Nucl. Phys. Proc. Suppl.} {\bf 45B}.

\bibitem {Appel} See Appelquist {\it et al\/}, ref.\cite{tHooft}

\bibitem{bando}
M. Bando, T. Kugo and K. Yamawaki 1988 {\it Phys. Rep.} {\bf 164}  217.

\bibitem{sannino}
T. Appelquist and F. Sannino 1999 {\it Phys. Rev.} {\bf D 59}  67702.

\bibitem{lattice} H. Wittig 2000 ``Strangeness in lttice QCD" e-print arXive: hep-ph/0008148, version 2, August; Alpha collaboration 2000  ``Effective chiral lagrangians and lattice QCD", e-print arXive: hep-lat/0006026, version 2, August; C.Allton 2000 ``Recent results from full lattice QCD", e-print archive: hep-ph/0005034, May.


\end{thebibliography}
\end{document}